\documentstyle[multicol,prl,aps,epsfig,color]{revtex}

\begin{document}
\draft

\title{Measurement of the Spin Asymmetry in the Photoproduction of
Pairs of High-$\boldmath p_T$ Hadrons at HERMES}

\author{
A.~Airapetian$^{33}$,
N.~Akopov$^{33}$,
M.~Amarian$^{25,28}$,
E.C.~Aschenauer$^{13,14,6}$,
H.~Avakian$^{10}$,
R.~Avakian$^{33}$,
A.~Avetissian$^{33}$,
B.~Bains$^{15}$,
C.~Baumgarten$^{23}$,
M.~Beckmann$^{12}$
% ...
\centerline {\it (The HERMES Collaboration)}
}

\address{
% ...
$^6$DESY, 15738 Zeuthen, Germany\\
% ...
$^{10}$Istituto Nazionale di Fisica Nucleare, Laboratori Nazionali di
Frascati, 00044 Frascati, Italy\\
% ...
$^{12}$Fakult\"at f\"ur Physik, Universit\"at Freiburg, 79104 Freiburg, Germany\\
$^{13}$Department of Subatomic and Radiation Physics, University of Gent,
9000 Gent, Belgium\\
$^{14}$Max-Planck-Institut f\"ur Kernphysik, 69029 Heidelberg, Germany\\
$^{15}$Department of Physics, University of Illinois, Urbana,
Illinois 61801, USA\\
% ...
$^{23}$Sektion Physik, Universit\"at M\"unchen, 85748 Garching, Germany\\
% ...
$^{25}$Nationaal Instituut voor Kernfysica en Hoge-Energiefysica (NIKHEF),
1009 DB Amsterdam, The Netherlands\\
% ...
$^{28}$Istituto Nazionale di Fisica Nucleare, Sezione Sanit\'a and
Physics Laboratory, Istituto Superiore di Sanit\'a, 00161 Roma, Italy\\
% ...
$^{33}$Yerevan Physics Institute, 375036, Yerevan, Armenia.
}
\date{\today}
\maketitle

\begin{abstract}
We present a measurement of the longitudinal spin asymmetry $A_{||}$ in
photoproduction of pairs of hadrons with high transverse 
momentum $p_T$.
Data were accumulated by the HERMES experiment using
a 27.5~GeV
polarized positron beam and 
a polarized hydrogen target internal to the HERA storage ring.
For $h^+h^-$ pairs with $p_T^{h_1}>1.5$~GeV/c and $p_T^{h_2}>1.0$~GeV/c, 
the measured asymmetry is $A_{||}=-0.28\pm 0.12~{\rm (stat.)}\pm
0.02~{\rm (syst.)}$.  This negative value is in contrast to the positive
asymmetries typically measured in deep inelastic scattering from
protons, and is interpreted
to arise from a positive gluon polarization.
\end{abstract}

\begin{multicols}{2}[]

From polarized deep inelastic lepton-nucleon scattering (DIS)
experiments, it has been inferred that the quark spins account for only a
fraction of the nucleon spin. One possible explanation
is a significant gluon polarization in the
nucleon. In principle, the
polarized gluon distribution $\Delta G(x_G)$ ($x_G$ is the fraction
of the nucleon momentum carried by the struck gluon) 
can be probed by a measurement of
the scaling violation of the polarized structure functions.  However,
the presently available data on polarized inclusive deep inelastic
scattering only poorly constrain $\Delta G(x_G)$, although there
is some indication for the integral to be positive
\cite{bib:ball,bib:e154,bib:smcfits}.
On the other hand, two theoretical calculations in the bag model
obtain different predictions for the sign of the integral of $\Delta G(x_G)$ 
\cite{bib:Jaffe,bib:Barone}.
Several recent experimental proposals have concentrated on ways 
to measure $\Delta G(x_G)$
directly\cite{bib:rhic,bib:compass,bib:hermescharm}.

One way to measure $\Delta G(x_G)$ directly is via the
photon gluon fusion process. Two useful experimental signatures
of this process are charm production and production of jets with 
high transverse momentum $p_T$.
In the former case, the large mass of the charm quark 
suppresses its production in the fragmentation process.
A similar argument applies to the production of jets: the transverse momentum 
produced in the fragmentation process is small and 
two back-to-back jets with sufficiently high
$p_T$ thus reflect the high $p_T$ of the quark and anti-quark produced
in the photon gluon fusion process.  Both charm production and high-$p_T$
jet production in DIS have resulted in direct measurements of the
unpolarized gluon structure function $G(x_G)$
\cite{bib:H1jpsi,bib:Zeusjpsi,bib:NMCcharm}.

At lower energy fixed target experiments, high-$p_T$ hadrons must serve 
in place of jets \cite{bib:bravar}. 
Several phenomenological studies of the potential of
high-$p_T$ meson photoproduction as a probe of $\Delta G(x_G)$ have
been performed\cite{bib:font,bib:carlson}.

In this Letter we present the first measurement of a spin asymmetry in
photoproduction of pairs of high-$p_T$ hadrons.
The data were collected in 1996 and 1997 by the HERMES experiment at
the HERA storage ring of the DESY laboratory.  Polarized positrons of
energy 27.5~GeV
were scattered off a polarized internal hydrogen gas
target. The beam polarization was continuously measured by Compton
back scattering and had an average value of $0.55 \pm 0.02$ 
\cite{bib:g1p,bib:longtrans}. The average target
polarization was
$0.86 \pm 0.04$ \cite{bib:g1p,bib:chrissy}.  
In both cases the quoted uncertainty is predominantly systematic.
The HERMES detector
\cite{bib:spectrometer} is a forward spectrometer that identifies
charged particles in the scattering angle range of 0.04 $< \theta <$
0.22 rad.  Particle identification (PID) is accomplished using an 
electromagnetic 
calorimeter, a scintillator hodoscope preceded by two radiation
lengths of lead, a transition radiation detector, and a
C$_4$F$_{10}$/N$_2$(70:30) gas threshold $\check{\rm C}$erenkov
counter.  A likelihood method, based on the empirical responses
of each of the four PID detectors, is used to discriminate between
positrons and hadrons. The
luminosity is measured in a pair of NaBi(WO$_4$)$_2$ electromagnetic
calorimeters that detect Bhabha-scattering from target electrons.

The longitudinal cross section asymmetry $A_{||}$ was determined 
using the formula:
\begin{equation}
  A_{\parallel}={{N^{\uparrow\downarrow}L^{\uparrow\uparrow} - N^{\uparrow\uparrow}L^{\uparrow\downarrow}}
    \over {N^{\uparrow\downarrow}L_P^{\uparrow\uparrow} + N^{\uparrow\uparrow}L_P^{\uparrow\downarrow}}}\:.
\end{equation}
Here $N^{\uparrow\uparrow}$($N^{\uparrow\downarrow}$) 
is the number of oppositely charged 
hadron pairs observed for target spin
parallel (anti-parallel) to the beam spin orientation.  The
luminosities for each target spin state are 
$L^{\uparrow\uparrow(\uparrow\downarrow)}$ and 
$L_P^{\uparrow\uparrow(\uparrow\downarrow)}$, the latter
being weighted by the product of the beam and target polarization values for
each spin state.

Events were selected that contained at least one positively charged
hadron $h^+$ and at least one negatively charged hadron $h^-$.  
The observation of the scattered positron was not required, in order
to include the very low $Q^2$ region which dominates the cross section.
Here $Q^2$ is the negative square of the 4-momentum of the virtual photon.
The highest momentum hadrons of each charge were required to have a
momentum above 4.5~GeV/c and a transverse momentum $p_T$ above 0.5~GeV/c.  
Here $p_T$ is defined as the momentum transverse to the
positron beam direction and is approximately equal to the momentum
transverse to the photon direction when $Q^2\approx 0$.  To
suppress contributions from vector meson resonances from the data
sample, a minimum value of the invariant mass of the two hadrons
(assuming both hadrons to be pions) $M(2\pi)>1.0$~GeV/c$^2$ was
imposed.  Additionally, both hadrons were required to originate from
the target region and to have a common vertex. 
A detailed account of the analysis may be found in
\cite{bib:jmartin}.

\begin{figure}[th]
\begin{center}
\epsfig{file=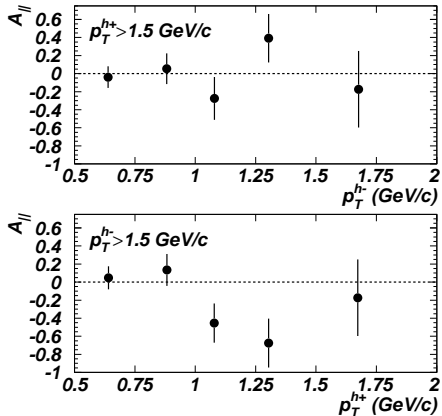,width=2.4in}
\begin{minipage}[r]{\linewidth}
\caption{\label{fig:neg}
  $A_{||}(p_T^{h^+},p_T^{h^-})$ for $p_T^{h^+}>1.5$~GeV/c (top)
  and for $p_T^{h^-}>1.5$~GeV/c (bottom). Note that the rightmost data point 
  is identical in both plots.}
\end{minipage}
\end{center}
\end{figure}

Fig.~\ref{fig:neg} presents the measured $A_{||}$ for the highest
values of transverse momenta accessible at HERMES; in the top (bottom)
panel
the positive
(negative) hadron was required to have a $p_T$ greater than 1.5~GeV/c
and the asymmetry $A_{||}$ is then plotted as a function of the $p_T$
of the hadron of opposite charge.  
The data suggest a more negative asymmetry when
the transverse momentum of the negative hadron is higher than that of
the positive hadron. Ignoring this charge asymmetry and
averaging over the five bins satisfying
the requirement $p_T^{h_1}>1.5$~GeV/c and $p_T^{h_2}>1.0$~GeV/c, 
a negative
asymmetry $A_{||}=-0.28\pm 0.12~{\rm (stat.)}\pm 0.02~{\rm (syst.)}$
is observed. (The symbol $h_1$ signifies the hadron with the higher $p_T$.) 
When the requirement $p_T^{h_1}>1.5$~GeV/c is not
enforced, the asymmetry is consistent with zero.  
The observed negative asymmetry is in contrast to the positive asymmetries
typically measured in deep inelastic scattering from protons.

A possible background to the observed asymmetry arises from coincident
detection of a negative hadron and the scattered positron, the latter
being misidentified as a positive hadron.  From studies of other
processes, the probability for positron/hadron misidentification has
been determined to be less than 0.2\%.  By comparing yields of
$h^+h^-$ pairs to those of 
$e^+h^-$ pairs detected in the final state, the background
arising from this misidentification has been estimated to be less than
$0.1\%$, for the kinematics selected by this analysis.
Other sources of background include high-$p_T$ particles  from charm decays. 
Contributions from both open charm and J/$\psi$ decays have been found to be  
negligible using the AROMA\cite{bib:aroma} Monte Carlo generator. 

The systematic uncertainty arising 
from the measurement of the beam and target polarizations is about 6\%
of $A_{||}$, much smaller than the statistical error and independent
of $p_T$. Resolution effects and alignment uncertainties were found to
be negligible. Electroweak radiative corrections are expected
to be very small compared to the statistical uncertainty.

The measured asymmetry was interpreted assuming that several
different processes
could contribute to the two-hadron 
cross section: lowest order 
deep inelastic scattering (containing no hard QCD vertex), 
interaction via the
hadronic structure of the photon -- described by the vector meson
dominance model (VMD) and by non-resonant hadronic ``anomalous''
photon structure, and the two first order QCD processes (termed ``direct'')
which describe the interaction of a pointlike photon.  These
are photon gluon fusion (PGF) and the QCD Compton effect (QCDC).

The contribution from lowest order DIS is suppressed by the requirement of high
$p_T$, and was confirmed to be negligible by a
simulation based on the LEPTO Monte Carlo generator \cite{bib:lepto}.
Contributions from VMD were assumed to have a negligible spin
asymmetry, and were thus treated as a dilution of the other asymmetries.
Finally, we neglect possible contributions from anomalous photon
structure, where the photon fluctuates into a non-resonant
$q\bar{q}$ pair which interacts via hard processes with the partons
inside the nucleon. This is supported by a model\cite{bib:Sch94}
that explains the excess of forward hadrons with high~$p_T$ 
observed in $\gamma p$ reactions at $70-90$~GeV,
 relative to those from $\pi p$ and
$Kp$ scattering\cite{bib:wa69}. At this energy, the model
prediction at high~$p_T$ is dominated by direct processes involving
hard coupling of the photon to the partons in the proton. At the 
lower energy of 
the present experiment, a negligible contribution from
anomalous photon structure is predicted by the model.

Under the assumptions described above, only two of 
the five possible spin asymmetries $A_i$ contribute significantly 
to the measured asymmetry: 
\begin{equation}
A_{||}\approx (A_{\rm PGF}f_{\rm PGF}+A_{\rm QCDC}f_{\rm QCDC})D
\end{equation}
where $f_i$ is the unpolarized fraction of events from subprocess $i$
($f_{\rm PGF}+f_{\rm QCDC}+f_{\rm VMD}=1$),
and $D$ is the virtual photon depolarization factor.  In the small
region of phase space selected by the present analysis, the $A_i$'s
may be approximated by the products of the hard subprocess asymmetries
and the quark and gluon polarizations.  The subprocess asymmetries
$\hat{a}_{\rm PGF}=\hat{a}(\gamma g \rightarrow q \bar{q})$ and
$\hat{a}_{\rm QCDC}=\hat{a}(\gamma q \rightarrow q g)$ are directly
calculable in leading order (LO) QCD \cite{bib:font}.  
For real photons and massless quarks, 
$\hat{a}_{\rm PGF} = -1$, while $\langle\hat{a}_{\rm QCDC}\rangle$ 
is about $+0.5$ (averaged over the kinematics selected by this
analysis) and is independent of the quark flavor.  The effective
quark polarization $\Delta q/q$ is computed as a suitably weighted combination
of
$\Delta u/u$ and $\Delta d/d$, which are known from
inclusive and semi-inclusive polarized DIS measurements
\cite{bib:deltaq,bib:smcsemi}.  The measured asymmetry can therefore
be expressed as follows:
\begin{equation}
A_{||} \approx \left( \hat{a}_{\rm PGF} 
\frac{\Delta G}{G}f_{\rm PGF} 
+ \hat{a}_{\rm QCDC}
\frac{\Delta q}{q}f_{\rm QCDC}\right) D , 
\label{eqn:apar}
\end{equation}
where the kinematic dependences have been suppressed for brevity.
This equation can be solved
for $\Delta G/G$ after appropriate averaging over the
selected kinematics.

The PYTHIA Monte Carlo generator \cite{bib:pythia} was used to 
provide a model for the data. An important parameter in
the simulation of the direct processes, the minimum transverse
momentum of the outgoing partons ($\hat{p}_T^{\rm min}$), was chosen
following Ref. \cite{bib:Sch94} to be 0.5~GeV/c.
The kinematic region used in the interpretation of the measurement
($p_T^{h_1}>1.5$~GeV/c and $p_T^{h_2}>0.8$~GeV/c)
was chosen so that the final results depend
only weakly on the choice of $\hat{p}_T^{\rm min}$.
The Lund fragmentation parameters used in the simulation have been
adjusted to fit the HERMES semi-inclusive 
hadron multiplicity data \cite{bib:tallini}.

\begin{figure}[th]
\begin{center}
\epsfig{file=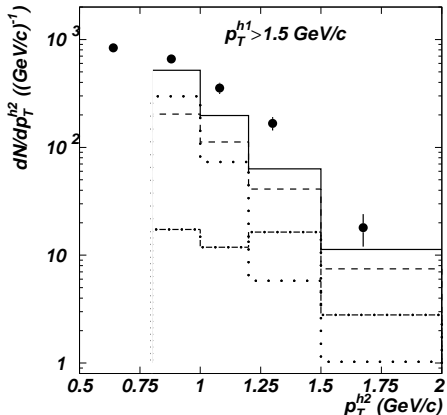,width=2.4in}
\begin{minipage}[r]{\linewidth}
\caption{\label{fig:xsect}
  Comparison of data (circles) and Monte Carlo simulation 
  (full histogram) for $dN/dp_T^{h_2}$ for $p_T^{h_1} > 1.5$~GeV/c.  
  The dashed, dashed-dotted and dotted lines represent the contributions 
  from the PGF, VMD and QCDC processes, respectively; the solid line represents
  their sum.}
\end{minipage}
\end{center}
\end{figure}

The normalized yield for the production of two high-$p_T$
hadrons is compared to the Monte Carlo simulation in
Fig.~\ref{fig:xsect}.  Here, the Weizs\"acker-Williams approximation
has been used to relate the photoproduction cross section simulated
by PYTHIA to the measured electroproduction cross section.  Also shown
in Fig.~\ref{fig:xsect} are the contributions from the three
subprocesses included in the simulation.  
The simulated yield has a $p_T$-dependence similar to that of the data,
but is significantly smaller in magnitude.
Good agreement is found for the distributions in 
other kinematic variables, such as the
azimuthal angle between the two hadrons
and $\Delta p_T = |\vec{p}_T^{h^-}|-|\vec{p}_T^{h^+}|$.  
As the simulation of the direct
QCD processes is restricted to leading order, the observed difference
in normalization might be due to contributions from higher-order QCD
processes and/or contributions from hard interactions of the hadronic
structure of the photon.
We also note that the agreement becomes much better if the default Lund
fragmentation parameters are used. However, the final result for $\Delta G/G$
is found to depend only weakly on the choice of fragmentation parameters.

\begin{figure}[th]
\begin{center}
\epsfig{file=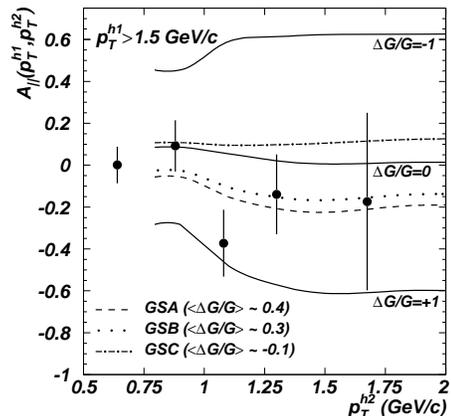,width=2.4in}
\begin{minipage}[r]{\linewidth}
\caption{\label{fig:bb}
  $A_{||}$ for high-$p_T$ hadron production measured at HERMES
  compared with Monte Carlo predictions for $\Delta G/G=\pm 1$
  (lower/upper solid curves), $\Delta G/G=0$ (middle solid curve), and
  the phenomenological LO QCD fits of Ref.~{\protect\cite{bib:gs}}
  (dashed, dotted, and dot-dashed curves).}
\end{minipage}
\end{center}
\end{figure}

In the same region of phase space where a negative asymmetry is
observed ($p_T^{h_1} > 1.5$~GeV/c and $p_T^{h_2} > 1.0$~GeV/c), the
simulated cross section is dominated by photon gluon fusion.  
The consequent sensitivity of
the measured asymmetry to the polarized gluon distribution is
demonstrated in Fig.~\ref{fig:bb}, where $A_{||}$ at high transverse
momenta (i.e. the average of the 
two panels of Fig.~\ref{fig:neg}) 
is compared with Monte Carlo predictions for different distributions
of $\Delta G/G$.

The Monte Carlo simulation was used to determine the quantities
necessary to relate the data to $\langle\Delta G/G\rangle$,
where the angle brackets indicate averaging over the kinematics of
the measurement.  
These quantities have been determined from PYTHIA to be
$\langle D\hat{a}_{\rm QCDC}\frac{\Delta q}{q}\rangle = 0.15$,
$\langle D\rangle = 0.93$, $\langle x_G\rangle = 0.17$, $\langle
Q^2\rangle = 0.06~{\rm (GeV/c)}^2$, and $\langle \hat{p}_T^2\rangle =
2.1$~(GeV/c)$^2$.  
The distribution $\Delta G(x_G)$ is probed principally in the range
$0.06 < x_G < 0.28$.
Note that the hard scale of this process is not
given by $Q^2$, but rather by $\hat{p}_T^2$, 
the square of the transverse momentum 
carried by each of the outgoing quarks.  

For the four values of $A_{||}$ at $p_T^{h_2} > 0.8$~GeV/c
presented in Fig.~\ref{fig:bb}, $\langle\Delta G/G\rangle$ 
was extracted according to equation (\ref{eqn:apar}). Since 
these four measurements 
probed essentially the same range of $x_G$, the results
for $\langle\Delta G/G\rangle$ were averaged.  Using the assumptions and model
parameters described above, a value for $\langle\Delta G/G\rangle$ 
was determined in LO QCD to be $0.41 \pm
0.18~{\rm (stat.)} \pm 0.03~{\rm (syst.)}$, where the systematic
uncertainty represents the experimental contribution only. 

\begin{figure}[th]
\begin{center}
\epsfig{file=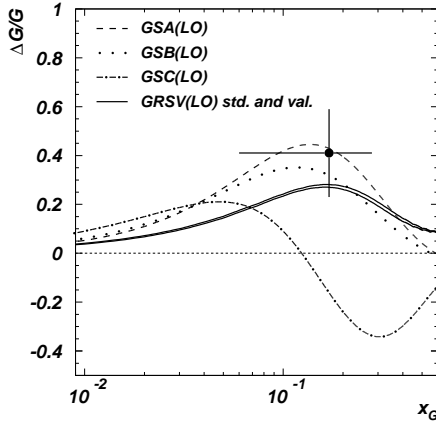,width=2.4in}
\begin{minipage}[r]{\linewidth}
\caption{\label{fig:dg}
  The extracted value of $\Delta G/G$ compared with phenomenological QCD
  fits to a subset of the world's data on $g_1^{\rm p,n}(x,Q^2)$.  The
  curves are from Refs.~{\protect\cite{bib:gs,bib:grsv}}, evaluated
  at a scale of 2~(GeV/c)$^2$.
  The indicated error on $\Delta G/G$ represents statistical and
  experimental systematic uncertainties only; no 
  theoretical uncertainty is included.}
\end{minipage}
\end{center}
\end{figure}

The extracted value of $\langle\Delta G/G\rangle$ 
is compared in Fig.~\ref{fig:dg}
with several phenomenological LO QCD fits of a subset of the world's
data on $g_1(x,Q^2)$ \cite{bib:gs,bib:grsv}.  The horizontal error bar
represents the standard deviation of the $x_G$ distribution for the
cited kinematical constraints on the produced hadrons, as given by the
Monte Carlo.  

In summary, a positive value for the gluon polarization has been
extracted from a measurement of the spin asymmetry in the 
photoproduction of pairs of hadrons at high $p_T$.
This interpretation of the observed negative asymmetry is based
on a model which takes into account leading order QCD processes
and VMD contributions to the cross section.
At the kinematics of this measurement, no spin-dependent analyses of
higher order QCD processes or contributions from
anomalous photon structure are presently
available; these processes have therefore been
neglected in the model presented here.  If such processes would be
important but have no significant spin asymmetry, the extracted value
of $\langle\Delta G/G\rangle$ 
would increase, but still differ from zero by 2.3$\sigma$.
To alter the principal conclusion of this analysis, i.e., that 
$\langle\Delta G/G\rangle$ at $\langle x_G\rangle =0.17$ is positive, a
significant contribution from a neglected process with a large
negative spin asymmetry would be needed.

We gratefully acknowledge the DESY management for its support, the DESY staff 
and the staffs of the collaborating institutions for their significant 
effort, and our funding agencies for financial support.

\end{multicols}
\end{document}